\documentclass{article}
\usepackage{eurosym}
\usepackage{amssymb}
\usepackage{amsmath}
\usepackage{cite}

\begin{document}

\title{Similarity solutions and Conservation laws for the
Bogoyavlensky-Konopelchenko Equation by Lie point symmetries}
\author{ Amlan K. Halder \and {\ \textit{Department of Mathematics, Pondicherry
University, Kalapet 605014, India}} \\
{\ \textit{Email: amlan91.res@pondiuni.edu.in}} \and Andronikos Paliathanasis\\
{\ \textit{Instituto de Ciencias F\'{\i}sicas y Matem\'{a}ticas, Universidad
Austral de Chile, Valdivia, Chile and}} \\
{\ \textit{Institute For Systems Science, Durban University of Technology }}\\
{\ \textit{PO Box 1334, Durban 4000, Republic of South Africa}} \\
{\ \textit{Email: anpaliat@phys.uoa.gr}} \\
\and P.G.L. Leach \\
{\ \textit{Institute For Systems Science, Durban University of Technology }}\\
{\ \textit{PO Box 1334, Durban 4000, Republic of South Africa and}} \\
{\ \textit{School of Mathematics, Statistics and Computer Science, University of KwaZulu-Natal }}\\
{\ \textit{Durban, Republic of South Africa}}\\
{\ \textit{Email: leachp@ukzn.ac.za}}}
\maketitle

\begin{abstract}
The $1+2$ dimensional Bogoyavlensky-Konopelchenko Equation is investigated
for its solution and conservation laws using the Lie point symmetry
analysis. In the recent past, certain work has been done describing the Lie point
symmetries for the equation and this work seems to be incomplete (Ray S (2017) Compt.
Math. Appl. \ 74, 1157). We obtained certain new symmetries and corresponding
conservation laws. The travelling-wave solution and some other
similarity solutions are studied.

Mathematics Subject Classiﬁcation (2010): 34A05, 34A34, 34C14, 22E60, 35B06, 35C05, 35C07.

\end{abstract}

\section{Introduction}

The $1+2$-dimensional Bogoyavlensky-Konopelchenko Equation is written as
\begin{equation}
q_{t}+\alpha q_{xxx}+\beta q_{xxy}+6\alpha qq_{x}+4\beta qq_{y}+4\beta
q_{x}\partial _{x}^{-1}q_{y}=0,  \label{1.1}
\end{equation}%
where $\alpha $ and $\beta $ are constants and for $\partial _{x}^{-1}q=u$,
(\ref{1.1}) is simplified as follows
\begin{equation}
u_{xt}+\alpha u_{xxxx}+\beta u_{xxxy}+6\alpha u_{xx}u_{x}+4\beta
u_{xy}u_{x}+4\beta u_{xx}u_{y}=0,  \label{1.2}
\end{equation}%
which is a two-dimensional extension of the Korteweg-de Vries (KdV) Equation.

In \cite{bk31,bk33} the authors have mentioned various applications of the
equation. Briefly, it is mentioned in \cite{bk31,bk05} that the equation
signifies the interaction between Riemann wave and long wave. Furthermore
certain particular cases were studied in \cite{bk04,bk03,bk05,bk32}.{}

Various method such as the inverse scattering method and Darboux transformation
are used to study the latter equation, for instance see \cite{bk03,bk04} and
references therein. Applications of the equations and the study of the
equation for various values of the arbitrary constants are mentioned in\cite%
{bk04,bk05}. In \cite{bk34} new position, negaton and complexiton solutions
are described for the BK Equation using the Darboux transformation. In \cite%
{bk02,bk02aa} the author considered the $1+2$-dimensional
Bogoyavlensky-Konopelchenko Equation as a system and obtained the Lie point
symmetries and used the condition of quasi self-adjointness to determine the
conservation laws. In \cite{bk06} some of the inaccuracies of \cite{bk02}
are mentioned and conservation laws using the more established Noether's
Theorem is presented.

In this work we study the Lie point symmetries of the \
Bogoyavlensky-Konopelchenko Equation and we apply them in order to study
certain possible reductions and determine closed-form and analytic solutions. The
theory of Lie symmetries of differential equations is the standard technique
for computation of solutions and describing the algebra for nonlinear
differential equations.

Our study provides new symmetries for the\ Bogoyavlensky-Konopelchenko
Equation which have not been reported before in the literature.  We we apply
these symmetries to determine new similarity solutions and also to determine
conservation laws \cite{bk37}.

Computation of conservation laws has been a core area of research for
decades. The standard technique is the Noether's Theorem for a system with a
Lagrangian. Later various researchers developed methods
\cite{bk37,bk39,bk39a,bk40,bk41} to compute the conservation laws for systems
without a Lagrangian by generalising Noether's Theorem. Among them the
most widely used and applicable method is the nonlinear self-adjointness
method developed by Ibragimov\cite{bk37,bk39}.

In this paper, for the determination of the conservation laws for equation (%
\ref{1.2}), we make use of the Ibragimov's method of nonlinearly self-adjointness. Our analysis extends the previous results of \cite{bk02} and shows
that the conservation laws for the $1+2$-dimensional
Bogoyavlensky-Konopelchenko Equation can be derived directly by the Lie
point symmetries. The plan of the paper it follows.

In Section \ref{lie2} the Lie point symmetries for the $1+2$-dimensional
Bogoyavlensky-Konopelchenko Equation are derived. In particular we find that
the equation admits five plus infinity Lie point symmetries. The application
of the Lie point symmetries is performed in Section \ref{lie3} in which
similarity solutions are derived for equation\ (\ref{1.2}). \ More
specifically we derive a family of travelling-wave solutions and some scaling
solutions. Ibragimov's method of nonlinear self-adjointness is discussed
in Section \ref{lie4}, in which also the generic formula and conditions for the
determination of conservation laws are given. The conservation laws (flows)
are presented in Section \ref{lie5} which generalise the results of \cite%
{bk02} and includes the results of \cite{bk06}. Finally in Section \ref%
{lie6}, we discuss our results and draw our conclusions. \hfill

\section{Lie point symmetry analysis}

\label{lie2}

For the convenience of the reader we briefly discuss the theory of Lie
symmetries of differential equations and the application of the differential
invariants for the construction of similarity solutions.

Let $\Phi $ describe the map of an one-parameter point transformation such
as $\Phi \left( u\left( x^{i}\right) \right) =u\left( x^{i}\right) \ $with
infinitesimal transformation%
\begin{eqnarray}
x^{i\prime } &=&x^{i}+\varepsilon \xi ^{i}\left( x^{i},u\right)
\label{sv.12} \\
u^{\prime } &=&u+\varepsilon \eta \left( x^{i},u\right)   \label{sv.13}
\end{eqnarray}%
and generator
\begin{equation}
\Gamma =\frac{\partial t^{\prime }}{\partial \varepsilon }\partial _{t}+%
\frac{\partial x^{\prime }}{\partial \varepsilon }\partial _{x}+\frac{%
\partial u^{A\prime }}{\partial \varepsilon }\partial _{u^{A}},  \label{sv.16}
\end{equation}
where~{$\varepsilon $ is the parameter of smallness} and $x^{i}=\left(
t,x,y\right) $. \ Assume that the function, $u\left( x^{i}\right), $ is a
solution of the partial differential equation $\mathcal{H}\left(
u^{A},u_{,t}^{A},u_{,x}^{A}...\right) =0$.  Then under the map, $\Phi $, the
function $u^{\prime }\left( x^{i\prime }\right) =\Phi \left( u\left(
x^{i}\right) \right) $ remains a solution for the differential equation if
and only if the differential equation is also invariant under the action of
the map $\Phi $, i.e., $\Phi \left( \mathcal{H}\left(
u^{A},u_{,t}^{A},u_{,x}^{A}...\right) \right) =0$.

When the latter is true, $\Gamma $ is called a Lie point symmetry for the
differential equation. Mathematically that is formulated with the following
condition
\begin{equation}
X^{\left[ n\right] }\left( \mathcal{H}\right) =0,  \label{sv.17}
\end{equation}%
in which $\Gamma ^{\left[ n\right] }$ describes the $n-$th
prolongation/extension of the symmetry vector in the jet-space of variables $%
\left\{ x^{i},u,u_{,i},u_{,ij},...\right\} $.

The importance of the existence of a Lie symmetry for a partial differential
equation is that from the associated Lagrange's system,%
\begin{equation}
\frac{dx^{i}}{\xi ^{i}}=\frac{du}{\eta },
\end{equation}%
zeroth-order invariants,~$U^{\left[ 0\right] }\left( x^{i},u\right) $ can be
determined which can be used to reduce the number of the independent
variables of the differential equation and lead to the construction of
similarity solutions.

As far as the Bogoyavlensky-Konopelchenko equation is concerned (\ref{1.2}) the
application of Lie's theory provides the generic symmetry vector field to
be
\begin{align}
\Gamma & =\begin{aligned}[t]&%
\bigg(A_{0}+(A_{1}-A_{2})t\bigg)\partial_t
+\nonumber\\
&\bigg(A_{3}+(A_{1}+A_{2})y+4t\left(A_{4}-\frac{(A_{1}+2A_{2})y}{4t\beta}\right)\beta%
\bigg)\partial_y +\nonumber\\
&\bigg(-A_{2}x+\frac{y(3A_{2}\alpha+(3A_{1}-2(A_{1}+2A_{2})+3A_{2})%
\alpha)}{\beta}+b(t)\bigg)\partial_x+\nonumber\\
&\bigg(A_{2}u+x\left(A_{4}-\frac{(A_{1}+2A_{2})y}{4t\beta}\right)+a(t)
+\frac{\frac{5}{2}A_{1}y^2\alpha+5A_{2}y^2\alpha-12A_{4}ty\alpha\beta+2ty%
\beta b'(t)}{8t\beta^2}\bigg)\partial_u,\nonumber\\ \end{aligned}  \label{2.1}
\\
&
\end{align}%
where $A_{0-4}$ are arbitrary constants and $a(t),b(t)$ are
arbitrary functions.

Hence the Lie symmetries corresponding to each of the arbitrary constants
and functions are
\begin{eqnarray}
\Gamma _{1a} &=&\partial _{t}  \notag  \label{2.2} \\
\Gamma _{2a} &=&t\partial _{t}+\frac{y\alpha }{\beta }\partial _{x}+\bigg(%
\frac{5y^{2}\alpha }{16t\beta ^{2}}-\frac{xy}{4t\beta }\bigg)\partial _{u}
\notag \\
\Gamma _{3a} &=&-t\partial _{t}-\bigg(x-\frac{2y\alpha }{\beta }\bigg)%
\partial _{x}-y\partial _{y}+\bigg(u+\frac{5y^{2}\alpha }{8t\beta ^{2}}-%
\frac{xy}{2t\beta }\bigg)\partial _{u}  \notag \\
\Gamma _{4a} &=&4t\beta \partial _{y}+\bigg(x-\frac{3y\alpha }{2\beta }\bigg)%
\partial _{u}  \notag \\
\Gamma _{5a} &=&\partial _{y}  \notag \\
\Gamma _{6a} &=&b(t)\partial _{x}+\bigg(a(t)+\frac{yb^{\prime }(t)}{\beta }%
\bigg)\partial _{u}.  \notag
\end{eqnarray}%
Below we continue with the application of the symmetry vectors and more
specifically with the determination of similarity solutions as also with the
construction of conservation laws.

\section{Similarity solutions}

\label{lie3}

We compute the travelling-wave solution for the Bogoyavlensky-Konopelchenko
Equation (\ref{1.2}) with respect to the generators $\Gamma _{6a}+\Gamma
_{5a}-c\Gamma _{1a},$ for $a(t)$ and $b(t)$ being constant with the similarity
variables defined as
\begin{eqnarray}
x+y-ct &=&s  \notag  \label{3.1} \\
u(t,x,y) &=&w(s).  \notag \\
&&
\end{eqnarray}%
\qquad \qquad \qquad

The fourth-order equation (\ref{1.2}) is reduced to the ordinary differential
equation,%
\begin{equation}
(\alpha +\beta )w^{\prime \prime \prime \prime }(s)+8\beta w^{\prime
}(s)w^{\prime \prime }(s)+6\alpha w^{\prime }(s)w^{\prime \prime
}(s)-cw^{\prime \prime }(s)=0.  \label{3.2}
\end{equation}%
The Lie point symmetries of this equation are
\begin{eqnarray}
\Gamma _{1b} &=&\partial _{s}  \notag  \label{3.3} \\
\Gamma _{2b} &=&\partial _{w}  \notag \\
\Gamma _{3b} &=&\partial _{s}+\bigg(\frac{cs}{7}-w\bigg)\partial _{w}.  \notag
\end{eqnarray}

The application of $\Gamma _{1b}$, reduces the fourth-order equation to one of
third order
\begin{equation}
v^{\prime \prime \prime }(r)=\frac{10 v^{\prime
}(r)v^{\prime \prime }(r)}{v(r)}-\frac{15v^{\prime 3}}{v(r)^{2}}+\bigg(\frac{cv(r)^{2}}{\alpha
+\beta }-\frac{14v(r)}{\alpha +\beta }\bigg)v^{\prime }(r),  \label{3.4}
\end{equation}%
where the variables are defined as
\begin{eqnarray}
r &=&w(s)  \notag  \label{3.5} \\
v(r) &=&\frac{1}{w^{\prime }(s)}.  \notag
\end{eqnarray}%
The third-order equation, (\ref{3.4}), has a lone point symmetry which is $\partial
_{r},~\ $because it is autonomous. The new similarity variables are
\begin{eqnarray}
h &=&v(r)  \notag  \label{3.6} \\
g(h) &=&\frac{1}{v(r)},  \notag
\end{eqnarray}%
and enables the reduction of (\ref{3.4}) to the second-order equation
\begin{equation}
g^{\prime \prime }(h)=\frac{3g^{\prime 2}}{g(h)}+\frac{10g^{\prime }(h)}{h}+%
\frac{(14h-h^{2}c)g(h)^{3}}{\alpha +\beta }+\frac{15g(h)}{h^{2}}  \label{3.7}
\end{equation}%
which is maximally symmetric and is easily integrated.

Now we consider reduction with respect to $\Gamma _{2b}$ with similarity
variables
\begin{eqnarray}
n &=&s  \notag  \label{3.8} \\
m(n) &=&w^{\prime }(s).  \notag
\end{eqnarray}%
The reduced third-order equation is
\begin{equation}
m^{\prime \prime \prime }(n)=m^{\prime }(n)\bigg(\frac{-14m(n)}{\alpha
+\beta }+\frac{c}{\alpha +\beta }\bigg).  \label{3.9}
\end{equation}%
This equation has two symmetries which are
\begin{eqnarray}
\Gamma _{1c} &=&\partial _{n}  \notag  \label{3.10} \\
\Gamma _{2c} &=&n\partial _{n}+\bigg(\frac{c}{7}-2m\bigg)\partial _{m}.
\notag
\end{eqnarray}%
The $\Gamma _{1c},$ with the variables $j=m(n)$ and $p(j)=\frac{1}{m^{\prime}(n)}$
leads to the second-order equation which is also maximally symmetric,
\begin{equation}
p^{\prime \prime }(j)=\frac{3p^{\prime 2}}{p(j)}-\frac{(c-14q)p(j)^{3}}{%
\alpha +\beta }.  \label{3.11}
\end{equation}

Interestingly $\Gamma _{2c}$ with the variables $l=\frac{(14m(n)-c)n^{2}}{14}$
 and $$k(l)=\frac{7}{n^{2}7nm^{\prime }(n)-c+14m(n)}$$ leads to the
second-order equation with no point symmetries, namely
\begin{equation}
k^{\prime \prime }(l)=\frac{3k^{\prime 2}}{k(l)}+9k(l)k^{\prime }(l)+\frac{%
(26\alpha +26\beta +14l)k(l)^{3}}{\alpha +\beta }-\frac{(24\alpha l+24\beta
l+28l^{2})k(l)^{4}}{\alpha +\beta }.  \label{3.12}
\end{equation}%
The reduction with respect to $\Gamma _{3b},$ leads to a third-order
equation with zero point symmetries.

The solutions provided by the set of equations (\ref{3.7}) and (\ref{3.11})
describe the travelling-wave solutions for the mother equation (\ref{1.2}).

\subsection{Further reductions based on other symmetries}

We study the equation (\ref{1.2}) with respect to $\Gamma _{2a}-\frac{%
\Gamma _{3a}}{2},$ which can be easily verified to be a symmetry. The
similarity variables are
\begin{eqnarray}
d &=&\frac{x}{y},  \notag  \label{3.13} \\
e &=&\frac{y^{3}}{t}\quad \mbox{\rm and}  \notag \\
p(d,e) &=&yu(t,x,y).  \notag
\end{eqnarray}%
The reduced PDE is
\begin{eqnarray}
d\beta p_{dddd}-3e\beta p_{ddde}-\alpha p_{dddd}+4\beta
p_{ddd}+c4^{2}p_{de}+4\beta pp_{dd}-12e\beta p_{e}p_{dd}\notag
\label{3.14} \\
-2p_{d}(6e\beta p_{de}+(3\alpha -4d\beta )p_{dd})+8\beta p_{d}^{2}&=&0.  \notag
\end{eqnarray}%
The Lie point symmetries are
\begin{eqnarray}
\Gamma _{1d} &=&e^{\frac{1}{3}}\partial _{p} \quad \mbox{and} \notag  \label{3.15} \\
\Gamma _{2d} &=&\frac{\partial _{d}}{e^{\frac{1}{3}}}+\frac{e^{\frac{2}{3}}}{%
12\beta }\partial _{p}.  \notag
\end{eqnarray}%
Reduction with respect to $\Gamma _{2d}$ leads to the solvable first-
order ode
\begin{equation}
r\beta ^{3}v(r)v^{\prime }(r)-r\beta v^{\prime }(r)-\beta v(r)+\frac{\beta
^{3}v(r)^{2}}{3}=0,  \label{3.16}
\end{equation}%
in which the variables are
\begin{eqnarray}
v(r) &=&\frac{12\beta p(e,d)}{ed}  \notag  \label{3.17} \\
r &=&e.  \notag
\end{eqnarray}

Further reduction based on $\Gamma _{4a}$, leads to the PDE
\begin{equation}
\alpha p_{dddd}+6\alpha p_{d}p_{dd}+\frac{p_{d}}{e}+p_{de}+\frac{dp_{dd}}{e}%
=0,  \label{3.18}
\end{equation}%
where the similarity variables are $d=x$, $t=e$ and $$u(t,x,y)=p(x,t)+\frac{%
(xy-\frac{3\alpha y^{2}}{4\beta })}{4t\beta }$$.

The Lie point symmetries for the latter equation (\ref{3.18}) are determined to be
\begin{eqnarray}
\Gamma _{1e} &=&\frac{d}{3}\partial _{d}+e\partial _{e}-\frac{p}{3}\partial
_{p}  \notag  \label{3.19} \\
\Gamma _{2e} &=&\partial _{e}+\frac{d^{2}}{12e^{2}\alpha }\partial _{p}
\notag \\
\Gamma _{3e} &=&\frac{4e^{\frac{3}{2}}}{3}\partial _{e}-\frac{2\sqrt{e}p}{3}%
\partial _{p} \notag \\
\Gamma _{4e} &=&\frac{d}{e}\partial _{p}-6\alpha \partial _{d} \quad \mbox{\rm and}  \notag \\
\Gamma _{5e} &=&e\partial _{d}.  \notag \\
&&
\end{eqnarray}

The reduction with respect to $\Gamma _{1e}$ leads to the fourth-order ode
\begin{eqnarray}
9 r\alpha [3 r v^{\prime \prime \prime \prime}(r)+8 v^{\prime \prime \prime}(r)-54 r^{2} \alpha v^{\prime \prime}(r)]+r[v^{\prime}(r)(r-24\alpha+162 r^{2} \alpha v^{\prime \prime}(r))]+ \label{3.20}\\
3r [2 (r + 12 \alpha) v^{\prime \prime}(r)-12 \alpha v(r)^{2}] + v(r) [r + 24 \alpha + 36 r \alpha v^{\prime}(r)]&=&0,\notag
\end{eqnarray}%
where $r=\frac{d^{3}}{e}$ and $p(d,e)=\frac{v(r)}{d}$. The sole Lie point
symmetry is $r^{\frac{1}{3}}\partial _{v}$. Accordingly the third-order ode
with zero Lie point symmetry is
\begin{equation}
m^{\prime \prime \prime }(n)=-\frac{4m^{\prime \prime }(n)}{n}-\left(\frac{2m(n)%
}{n^{\frac{2}{3}}}+\frac{6n^{\frac{8}{3}}+180n^{\frac{5}{3}}\alpha }{n^{%
\frac{11}{3}}\alpha}\right)m^{\prime }(n)-\frac{4m(n)^{2}}{3n^{\frac{5}{3}}}-%
\frac{5m(n)}{81n^{2}\alpha }.  \label{3.21}
\end{equation}

Similarly with respect to $\Gamma _{2e}$, the reduced ODE is
\begin{equation}
\alpha v^{\prime \prime \prime \prime }(r)+6\alpha v^{\prime }(r)v^{\prime
\prime }(r)=0,  \label{3.22}
\end{equation}%
where the similarity variables are $d=r$ and $p(d,e)=v(r)-\frac{d^{2}}{%
12e\alpha }$. The Lie point symmetries are
\begin{eqnarray}
\Gamma _{1f} &=&\partial _{r}  \notag  \label{3.23} \\
\Gamma _{2f} &=&\partial _{v} \quad \mbox{\rm and} \notag \\
\Gamma _{3f} &=&r\partial _{r}-v\partial _{v}.  \notag \\
&&
\end{eqnarray}

The reduction with respect to $\Gamma _{1f}$, with similarity variables $%
n=v(r)$ and $m(n)=\frac{1}{v(r)}$, leads to a third-order equation with
two symmetries.   It is
\begin{equation}
m^{\prime \prime \prime }(n)=-6m^{\prime }(n)m(n)+\frac{10m^{\prime
}(n)m^{\prime \prime }(n)}{m(n)}-\frac{15m^{\prime 3}}{m(n)^{2}}
\label{3.24}
\end{equation}%
and the two symmetries are $\partial _{n}$ and $n\partial _{n}-2m\partial
_{m}$. With respect to $\partial _{n}$ the equation is reduced to a maximally
symmetric second-order equation, which is
\begin{equation}
b^{\prime \prime }+\frac{3b^{\prime 2}}{b(a)}+\frac{10b^{\prime }(a)}{a}+%
\frac{15b(a)}{a^{2}}=0.  \label{3.25}
\end{equation}%
With respect to the other symmetry, (\ref{3.24}) is reduced to a second-order equation
with no Lie point symmetries, namely,
\begin{equation*}
b^{\prime \prime }(a)=\frac{3b^{\prime 2}}{b(a)}+\left(\frac{10}{a}%
-11b(a)\right)b^{\prime }(a)+\frac{15b(a)}{a^{2}}-\frac{40b(a)^{2}}{a}+\frac{
(6a^{3}+46a^{2})b(a)^{3}}{a^{2}}-\frac{(12a^{4}+24a^{3})b(a)^{4}}{a^{2}}.
\end{equation*}

Reduction with respect to $\Gamma _{2f}$ leads to the equation
\begin{equation}
m^{\prime \prime \prime }(n)=-6m^{\prime }(n)m(n),  \label{3.27}
\end{equation}%
where $n=r$ and $m(n)=v^{\prime }(r)$. The symmetries are $\partial _{n}$
and $n\partial _{n}-2m\partial _{m}$. The $\partial _{n}$ symmetry reduces
it to
\begin{equation}
b+\frac{3b^{\prime 2}}{b(a)}=0  \label{3.28}
\end{equation}%
which is linearisable. As in the case of $\Gamma _{1f}$ the $n\partial
_{n}-2m\partial _{m}$ symmetry leads to a equation with no Lie point
symmetries.

The last possible reduction with respect to $\Gamma_{3f}$ leads to a
third-order equation with no Lie point symmetries. The other possible
reductions with respect to other symmetries do not have a concrete result
and hence we omit those.

However, the interesting part of our analysis is the determination of the
conservation laws for the Bogoyavlensky-Konopelchenko equation with the use of
Ibragimov's method of Nonlinear Self-adjointness.

\section{Nonlinear self-adjointness and conservation laws}

\label{lie4}

In this Section we mention the preliminaries of Ibragimov's method of
Nonlinear Self-adjointness. Let
\begin{equation}
E(x,u,u_{i},u_{ij},..)=0  \label{4.1}
\end{equation}%
be a scalar PDE, where $x=(x^{i},x^{j},x^{k},..)$ are the independent
variables and $u_{i}$, $u_{ij}$ are defined as
$u_{i}=\frac{\partial _{u}}{\partial _{x^{i}}}$ , $u_{ij}=\frac{\partial _{u}}{\partial _{x^{i}x^{j}}}.$
Similarly we can define $u_{ijk}$ and other terms. The adjoint system to
the scalar PDE can be defined as
\begin{equation*}
E^{\ast }(x,u,u_{i},u_{ij},..)=\frac{\delta {\mathit{L}}}{\delta u},
\end{equation*}%
where $\mathit{L}$ can be defined as
\begin{equation*}
\mathit{L}=v(x)E(x,u,u_{i},u_{ij},..).
\end{equation*}%
$v(x)$ is the new dependent variable and $x$ is defined as $%
x=(x^{i},x^{j},x^{k},..)$. The variational derivative $\frac{\delta }{\delta
u}$ is
\begin{equation}
\frac{\delta }{\delta u}=\frac{\partial }{\partial _{u}}+\sum_{s=1}^{\infty
}(-1)^{s}D_{i_{1}}D_{i_{2}}...D_{i_{s}}\frac{\partial }{\partial
_{u_{i_{1}i_{2}...i_{s}}}}.
\end{equation}%
Now to verify the nonlinear self-adjointness, the substitution of $v=\phi
(x,u)$, where $\phi (x,u)\neq 0$, to the adjoint equation of the scalar PDE
must satisfy for all solutions, $u$, of equation (\ref{4.1}). In other words
\begin{equation}
E^{\ast }(x,u,u_{i},u_{ij},..)=\lambda E(x,u,u_{i},u_{ij},..),
\end{equation}%
where $\lambda $ is the undetermined coefficient.

\subsection{Conservation laws}

Let the scalar PDE admit the following generators of the infinitesimal
transformation
\begin{equation}
V=\xi ^{i}(x,u,u_{i},..)\frac{\partial }{\partial _{x}^{i}}+\eta
(x,u,u_{i},...)\frac{\partial }{\partial _{u}}.
\end{equation}%
Then the scalar PDE and its adjoint equation as defined above admits the
conservation law
\begin{equation*}
C^{i}=\xi ^{i}\mathit{L}+W[\frac{\partial {\mathit{L}}}{\partial _{u_{i}}}%
-D_{j}(\frac{\partial {\mathit{L}}}{\partial _{u_{ij}}})+D_{j}D_{k}(\frac{%
\partial {\mathit{L}}}{\partial _{u_{ijk}}})-...]+D_{j}(W)[\frac{\partial {%
\mathit{L}}}{\partial _{u_{ij}}}-D_{k}(\frac{\partial {\mathit{L}}}{\partial
_{u_{ij}}})+...]+D_{j}D_{k}(W)[\frac{\partial {\mathit{L}}}{\partial
_{u_{ijk}}}-...],
\end{equation*}%
where $W=\eta -\xi ^{i}u_{i}$.

\subsection{Condition for nonlinear self-adjointness}

We employ Ibragimov's method of Nonlinear Self-adjointness to construct the
conservation laws for (\ref{1.2}). We firstly prove that equation (\ref%
{1.2}) is indeed nonlinearly self-adjoint by considering the Lagrangian as
\begin{equation}
\mathit{L}=v(t,x,y)(u_{xt}+\alpha u_{xxxx}+\beta u_{xxxy}+6\alpha
u_{xx}u_{x}+4\beta u_{xy}u_{x}+4\beta u_{xx}u_{y}),  \label{5.1}
\end{equation}%
where v(t,x,y) is the new dependent variable. To verify the nonlinear
self-adjointness, the substitution of $v=\phi (t,x,y,u)$ into the adjoint
equation of (\ref{1.2}) must satisfy for all solutions, $u$, of equation (\ref%
{1.2}). The adjoint equation for (\ref{1.2}) is
\begin{eqnarray*}
\frac{\delta {\mathit{L}}}{\delta u} &=&6\alpha u_{x}v_{xx}+4\beta
u_{y}v_{xx}+6\alpha u-{xx}v_{x}+4\beta u_{yy}+v_{xt}+ \\
&&4\beta u_{x}v_{xy}+4\beta v_{x}u_{xy}+\alpha v_{xxxx}+\beta v_{xxxy}.
\end{eqnarray*}

The corresponding derivative terms are
\begin{eqnarray}
v_{x} &=&\phi _{x}+\phi _{u}u_{x},  \notag  \label{5.2} \\
v_{y} &=&\phi _{y}+\phi _{u}u_{y},  \notag \\
v_{xx} &=&\phi _{xx}+u_{x}(\phi _{uu}u_{x})+\phi _{u}u_{xx},  \notag \\
v_{xy} &=&\phi _{xy}+u_{x}(\phi _{uu}u_{y})+\phi _{u}u_{xy},  \notag \\
v_{xt} &=&\phi _{xt}+u_{x}(\phi _{uu}u_{t})+\phi _{u}u_{xt},  \notag \\
v_{xxx} &=&\phi _{xxx}+u_{x}^{2}(\phi _{uuu}u_{x})+2\phi
_{uu}u_{x}u_{xx}+u_{xx}\phi _{u}u_{x}+\phi _{u}u_{xxx},  \notag \\
v_{xxxx} &=&\phi _{xxxx}+u_{x}^{3}\phi _{uuuu}u_{x}+3\phi
_{uuu}u_{x}^{2}u_{xx}+...  \quad \mbox{\rm and}\notag \\
v_{xxxy} &=&\phi _{xxxy}+...\,.
\end{eqnarray}%
We substitute the values of $v$ into the adjoint equation and look for the
possible values for $\phi (t,x,y,u)$.
\begin{eqnarray}
\left(
\begin{array}{c}
6\alpha u_{x}v_{xx}+4\beta u_{y}v_{xx}+6\alpha u-{xx}v_{x}+4\beta u_{yy}+ \\
+v_{xt}+4\beta u_{x}v_{xy}+4\beta v_{x}u_{xy}+\alpha v_{xxxx}+\beta v_{xxxy}%
\end{array}%
\right) &=&\lambda (u_{xt}+\alpha u_{xxxx}+\beta u_{xxxy}  \notag
\label{5.3} \\
&&+6\alpha u_{xx}u_{x}+4\beta u_{xy}u_{x}+4\beta u_{xx}u_{y}).
\end{eqnarray}%
Substituting (\ref{5.2}) into (\ref{5.3}), we obtain the undetermined
coefficient $\lambda =0$ and $\phi (t,x,y,u)=A_{5},$ where $A_{5}$ is a
constant. Hence equation (\ref{1.2}) satisfies the Nonlinearly Self-adjointness condition.

\section{The conservation laws}

\label{lie5}

Corresponding to each of the symmetries for equation (\ref{1.2}), we obtain the fluxes
for the corresponding density. These conservation laws extend and complete
previous results in the literature.

For $\Gamma _{1a}$ the conserved fluxes are
\begin{eqnarray}
c^{t} &=&\phi (t,x,y,u)(u_{xt}+\alpha u_{xxxx}+\beta u_{xxxy}+6\alpha
u_{xx}u_{x}+4\beta u_{xy}u_{x}+4\beta u_{xx}u_{y}),  \notag \\
c^{x} &=&-\phi (t,x,y,u)u_{xt}(6\alpha u_{x}+4\beta u_{y}) \quad \mbox{\rm and} \notag \\
c^{y} &=&-\phi (t,x,y,u)u_{t}4\beta u_{xx}.  \notag
\end{eqnarray}

For $\Gamma _{2a}$ the conserved fluxes are
\begin{align}
c^{t}& =\begin{aligned}[t] &t\phi(t,x,y,u)(u_{xt}+ \alpha u_{xxxx} + \beta
u_{xxxy} + 6 \alpha u_{xx} u_{x} + &4 \beta u_{xy}u_{x} +4 \beta
u_{xx}u_{y}),\nonumber\\ \end{aligned} \\
c^{x}& =\begin{aligned}[t] &\frac{y\alpha}{\beta}\phi(t,x,y,u)(u_{xt}+
\alpha u_{xxxx} + \beta u_{xxxy} + 6 \alpha u_{xx} u_{x} + 4 \beta
u_{xy}u_{x} +4 \beta u_{xx}u_{y})-\nonumber\\
&\phi(t,x,y,u)\left(\frac{y}{4t\beta}+t
u_{xt}+\frac{y\alpha}{\beta}u_{xx}\right)(6\alpha u_{x}+4\beta u_{y}) \quad \mbox{\rm and}\nonumber\\
\end{aligned} \\
c^{y}& =\begin{aligned}[t] &4\phi(t,x,y,u)\beta
u_{xx}\left(\frac{5y^2\alpha}{16t\beta^2}-\frac{xy}{4t\beta}-tu_{t}-\frac{y%
\alpha}{\beta}u_{x}\right).\nonumber\\ \end{aligned}
\end{align}

For $\Gamma _{3a}$ the conserved fluxes are
\begin{align}
c^{t}& =\begin{aligned}[t] &-t\phi(t,x,y,u)(u_{xt}+ \alpha u_{xxxx} + \beta
u_{xxxy} + 6 \alpha u_{xx} u_{x} +\nonumber\\ &4 \beta u_{xy}u_{x} +4 \beta
u_{xx}u_{y}),\nonumber\\ \end{aligned} \\
c^{x}& =\begin{aligned}[t] &\left(\frac{2y\alpha}{\beta}-x\right)\phi(t,x,y,u)(u_{xt}+
\alpha u_{xxxx} + \beta u_{xxxy} + 6 \alpha u_{xx} u_{x}+\nonumber\\ &4
\beta u_{xy}u_{x} +4 \beta u_{xx}u_{y})+
\left(yu_{xy}+tu_{xt}+xu_{xx}+2u_{x}-\frac{y}{2t\beta}\right)\phi(t,x,y,u)(6\alpha
u_{x}+4 \beta u_{y}) \quad \mbox{\rm and}\nonumber\\ \end{aligned} \\
c^{y}& =\begin{aligned}[t] &-y\phi(t,x,y,u)(u_{xt}+ \alpha u_{xxxx} + \beta
u_{xxxy} + 6 \alpha u_{xx} u_{x}+\nonumber\\ &4 \beta u_{xy}u_{x} +4 \beta
u_{xx}u_{y})+4\beta
\phi(t,x,y,u)u_{xx}(u+\frac{5y^2\alpha}{8t\beta^2}-\frac{xy}{2t\beta}+%
\nonumber\\ &tu_{t}+xu_{x}-\frac{2y\alpha}{\beta}u_{x}+yu_{y}).\nonumber\\
\end{aligned}
\end{align}

For $\Gamma _{4a}$ the nonzero conserved fluxes are
\begin{align}
c^{x}& =\begin{aligned}[t] &\phi(t,x,y,u)(1-4t\beta u_{xy})(6\alpha u_{x}+4
\beta u_{y}) \quad \mbox{\rm and}\nonumber\\ \end{aligned} \\
c^{y}& =\begin{aligned}[t] &4t\beta \phi(t,x,y,u)(u_{xt}+ \alpha u_{xxxx} +
\beta u_{xxxy} + 6 \alpha u_{xx} u_{x} +4 \beta u_{xy}u_{x} +4 \beta
u_{xx}u_{y})+(x-\frac{3y\alpha}{2\beta}-4t\beta u_{y})4\beta
\phi(t,x,y,u)u_{xx} \nonumber\\ \end{aligned}
\end{align}

For $\Gamma _{5a}$ the nonzero conserved fluxes are
\begin{align}
c^{x}& =\begin{aligned}[t] &-u_{xy}\phi(t,x,y,u)(6\alpha u_{x}+4 \beta
u_{y}) \quad \mbox{\rm and}\nonumber\\ \end{aligned} \\
c^{y}& =\begin{aligned}[t] &\phi(t,x,y,u)(u_{xt}+ \alpha u_{xxxx} + \beta
u_{xxxy} + 6 \alpha u_{xx} u_{x} + 4 \beta u_{xy}u_{x} +4 \beta
u_{xx}u_{y})-u-{y}\phi(t,x,y,u)4\beta u_{xx}.\nonumber\\ \end{aligned}
\end{align}

For $\Gamma _{6a}$ the nonzero conserved fluxes are
\begin{align}
c^{x}& =\begin{aligned}[t] &b(t)\phi(t,x,y,u)(u_{xt}+ \alpha u_{xxxx} +
\beta u_{xxxy} + 6 \alpha u_{xx} u_{x} + 4 \beta u_{xy}u_{x} +4 \beta
u_{xx}u_{y})-b(t)\phi(t,x,y,u)u_{xx}(6\alpha u_{x}+4 \beta u_{y})\nonumber\\
\end{aligned} \\
c^{y}& =\begin{aligned}[t]
&4\beta\phi(t,x,y,u)u_{xx}(a(t)+\frac{yb'(t)}{4\beta}-b(t)u_{x})\nonumber\\
\end{aligned}
\end{align}

For $\Gamma _{7a}$, the nonzero conserved vectors are
\begin{align}
c^{t}& =\begin{aligned}[t] &\frac{-3t}{2}\phi(t,x,y,u)(u_{xt}+ \alpha
u_{xxxx} + \beta u_{xxxy} + 6 \alpha u_{xx} u_{x} + 4 \beta u_{xy}u_{x} +4
\beta u_{xx}u_{y}),\nonumber\\ \end{aligned} \\
c^{x}& =\begin{aligned}[t] &-\frac{x}{2}\phi(t,x,y,u)(u_{xt}+ \alpha
u_{xxxx} + \beta u_{xxxy} + 6 \alpha u_{xx} u_{x} + 4 \beta u_{xy}u_{x} +4
\beta u_{xx}u_{y})+\nonumber\\ &\left(u_{x}+\frac{x u_{xx}}{2}+\frac{y
u_{xy}}{2}+\frac{3t u_{xt}}{2}\right)\phi(t,x,y,u)(6\alpha u_{x}+4\beta
u_{y}) \quad \mbox{\rm and} \nonumber\\ \end{aligned} \\
c^{y}& =\begin{aligned}[t] &-\frac{y}{2}\phi(t,x,y,u)(u_{xt}+ \alpha
u_{xxxx} + \beta u_{xxxy} + 6 \alpha u_{xx} u_{x} + 4 \beta u_{xy}u_{x} +4
\beta u_{xx}u_{y})+\nonumber\\
&4\beta\phi(t,x,y,u)u_{xx}\left(\frac{u}{2}+\frac{x u_{x}}{2}+\frac{y
u_{y}}{2}+\frac{3t u_{t}}{2}\right).\nonumber\\ \end{aligned}
\end{align}

\section{Conclusions}

\label{lie6}

In this work we applied symmetry analysis to determine similarity
solutions and conservation laws for the $1+2$-dimensional
Bogoyavlensky-Konopelchenko Equation. More specifically we considered the
Lie point symmetries and we studied the reduction process in order to find
similiarity solutions and also determined conservation laws.

The application of Lie's theory provides us that equation (\ref{1.2}) admits
five plus infinity Lie symmetries. The infinity symmetries are directed
related with the linearity of the Bogoyavlensky-Konopelchenko Equation. We
apply the five finite Lie symmetries to reduce the differential equation
to an ordinary differential equation, while for the latter one we apply
again the Lie symmetries to reduce the order of the ordinary differential
equation and consequently determine similarity solutions, or reduce the
equation to well-known integrable differential equations. This analysis
extends previously published results in the literature \cite{bk02,bk02aa}. Our future work is to study the integrability 
of the reduced ODEs with zero point symmetries using Singularity Analysis.

Furthermore we apply Ibragimov's approach to determine conservation
laws by constructing self-adjoint operators from the admitted Lie point
symmetries.

\section{Acknowledgements}
AKH expresses grateful thanks to UGC (India), NFSC, Award No.
F1-17.1/201718/RGNF-2017-18-SC-ORI-39488 for financial support and to Late Prof. K.M.Tamizhmani for the discussions which formed the base of the work. AP
acknowledges the financial support of FONDECYT grant no. 3160121. PGLL
Thanks the Durban University of Technology, the University of KwaZulu-Natal
and the National Research Foundation of South Africa for support.





\end{document}